\documentclass{aip-cp}
\usepackage{cite}
\usepackage{tabularx}
\usepackage{rotating}
\usepackage{graphicx}
\usepackage{hyperref}
\usepackage{color}
\usepackage{epstopdf}
\usepackage{cleveref}
\usepackage[svgnames]{xcolor}
\usepackage{enumerate}
\usepackage{braket}
\usepackage{cite}
\hypersetup{hidelinks,colorlinks=true,allcolors=DarkBlue}
\usepackage{indentfirst}
\usepackage{amsfonts,amssymb}
\usepackage{subfigure}
\usepackage{caption}
\usepackage{braket}
\usepackage{siunitx}
\usepackage{dcolumn}
\hypersetup{hidelinks,colorlinks=true,allcolors=DarkBlue}

\begin{document}

\title{SPHINCS$^+$ post-quantum digital signature scheme \\ with Streebog hash function}

\author[aff1,aff2]{E.O. Kiktenko}\corresp[cor1]{Corresponding author: e.kiktenko@rqc.ru (E.O.K.)}
\author[aff1]{A.A. Bulychev }
\author[aff1,aff2]{P.A. Karagodin}
\author[aff1,aff2]{N.O. Pozhar}
\author[aff1,aff2]{M.N. Anufriev}
\author[aff1,aff2]{A.K. Fedorov}\corresp[cor2]{Corresponding author: akf@rqc.ru (A.K.F.)}

\affil[aff1]{Russian Quantum Center, Skolkovo, Moscow 143025, Russia}
\affil[aff2]{QApp, Skolkovo, Moscow 143025, Russia}

\maketitle

\begin{abstract}
Many commonly used public-key cryptosystems will become insecure once a scalable quantum computer is built.
New cryptographic schemes that can guarantee protection against attacks with quantum computers, so-called post-quantum algorithms, have emerged in recent decades.
One of the most promising candidates for a post-quantum signature scheme is SPHINCS$^+$, which is based on cryptographic hash functions.
In this contribution, we analyze the use of the new Russian standardized hash function, known as Streebog, for the implementation of the SPHINCS$^+$ signature scheme.
We provide a performance comparison with SHA-256-based instantiation and give benchmarks for various sets of parameters.
\end{abstract}

\section{Introduction}

Public-key cryptography is a cornerstone of internet security. 
Quantum computers possess a threat to the widely deployed public-key cryptography schemes, whose security is based on the computational complexity of certain tasks, such as integer factorization and discrete logarithm.
Shor's quantum algorithm~\cite{Shor1997} and variational quantum factoring~\cite{Anschuetz2018} would allow one to solve these tasks with a significant boost~\cite{ETSI}. 
Quantum computers have less of an effect on symmetric cryptography since Shor's algorithm does not apply for their cryptoanalysis. 
Nevertheless, Grover's algorithm~\cite{Grover1996} would allow quantum computers a quadratic speedup in brute force attacks.
Thus, the current goal is to develop cryptographic systems that are secure against both classical and quantum attacks, before large-scale quantum computers arrive. 

Fortunately, not all public-key cryptosystems are vulnerable to attacks with quantum computers~\cite{Bernstein2009}.
Several cryptosystems, which strive to remain secure under the assumption that the attacker has a large-scale quantum computer, have been suggested~\cite{Bernstein2017}.
These schemes are in the scope of so-called post-quantum cryptography. 
Existing proposals for post-quantum cryptography include code-based and lattice-based schemes for encryption and digital signatures as well as signature schemes based on hash functions. 

Hash-based digital signatures are built upon cryptographic hash functions, which are well-known tools in the modern cryptography. 
Such schemes attract significant attention since their security can be reduced to the properties of the chosen hashing primitive. Another benefit of hash-based signature schemes is their flexibility as they can be used with any secure hashing function, and so if a flaw is discovered in a secure hashing function, 
a hash-based signature scheme just needs to switch to a new and secure hash function to remain effective.
The most advanced version of hash-based digital signatures is SPHINCS$^+$~\cite{sphincs-pls}, which is a modification of the previously suggested SPHINCS scheme employing the Merkle hyper-tree~\cite{Bernstein2015}.
Another option is the Gravity-SPHINCS scheme~\cite{GravitySPHINCS}, whose primary innovation is the authentication scheme update. 
Nevertheless, SPHINCS$^+$ requires fewer security assumptions.  

In the present work, we consider the use of the new Russian standardized hash function, known as Streebog (GOST R 34.11-2012), for the implementation of the SPHINCS$^+$ signature scheme.
The Streebog hash function, described in RFC 6986~\cite{RFC}, is of Merkle-Damgard-type function, which makes it suitable for the installation in the SPHINCS$^+$ scheme. 
We provide a performance comparison with SHA-256-based instantiation and give benchmarks for various sets of parameters.

\begin{figure}[h]
\label{fig-sphincs}
\begin{minipage}[h]{1\linewidth}
\centerline{\includegraphics[width=0.7\columnwidth]{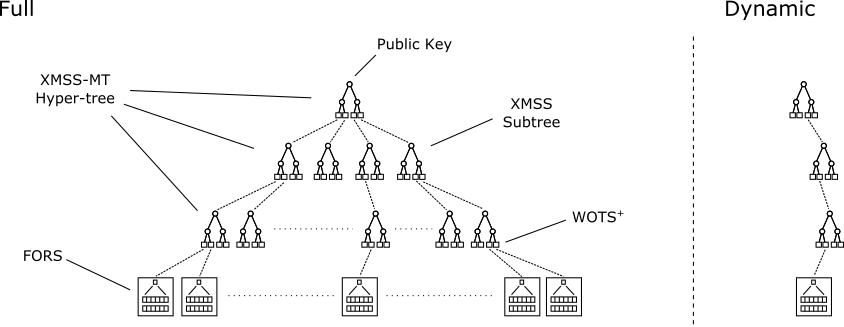}\label{fig-sphincs}}
\centerline{\mbox{\textbf{Fig.~1}. SPHINCS$^+$ signatures scheme construction:}}
\centerline{Full scheme of the signatures construction (left) and the dynamic representation (right).}
\end{minipage}
\end{figure}

\section{SPHINCS$^+$ instantiation with Streebog hash function}

The main goal of this paper is to analyze the use of the Russian standardized hash functions for the SPHINCS$^+$ signature scheme.
To this end, we start from brief overview of SPHINCS$^+$ and then demonstrate the results of its work with the Russian standardized hash function Streebog, specified in GOST R 34.11-2012 and described in RFC 6986~\cite{RFC}.

\begin{figure}[h]
\begin{minipage}[h]{1\linewidth}
\centerline{\includegraphics[width=0.7\columnwidth]{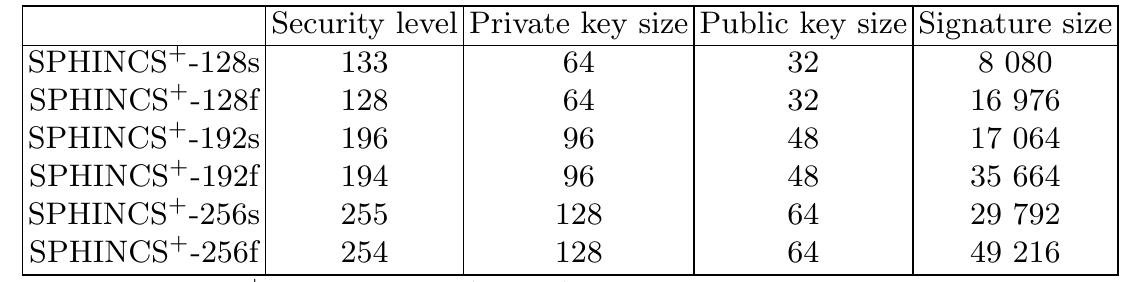}}
\centerline{\mbox{\textbf{Table.~I}. SPHINCS$^+$ security level (in bits), private key, public key, and signature sizes (in bytes)}}
\centerline{for different specifications of the signature scheme~\cite{sphincs-pls}.}
\label{table-sizes}
\end{minipage}
\end{figure}

\begin{figure}[h]
\begin{minipage}[h]{1\linewidth}
\centerline{\includegraphics[width=0.7\columnwidth]{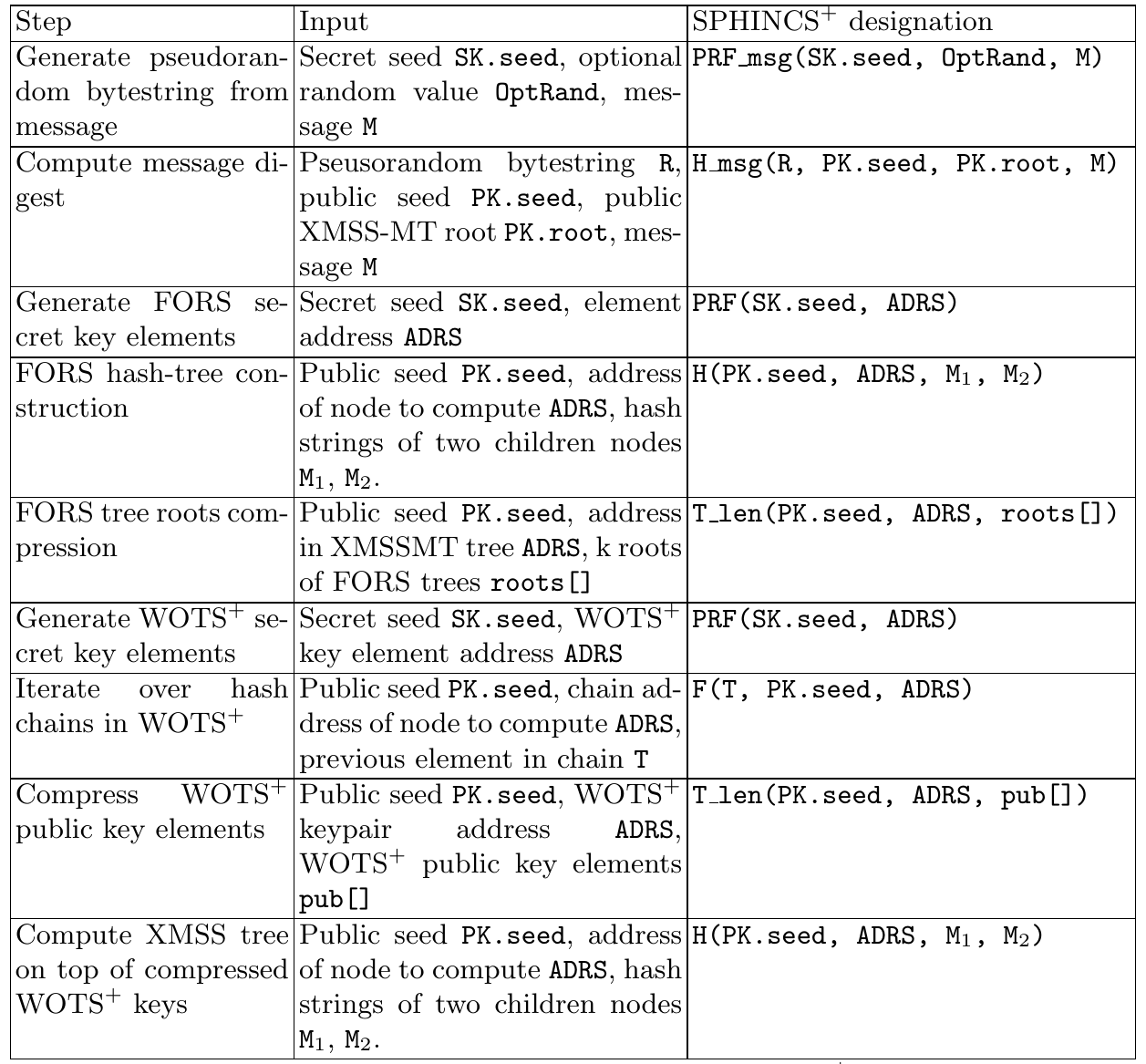}}
\centerline{\mbox{\textbf{Table.~II}. Hash functions usage in SPHINCS$^+$.}}
\label{hash-usage}
\end{minipage}
\end{figure}

\begin{figure}[h]
\begin{minipage}[h]{1\linewidth}
\centerline{\includegraphics[width=0.7\columnwidth]{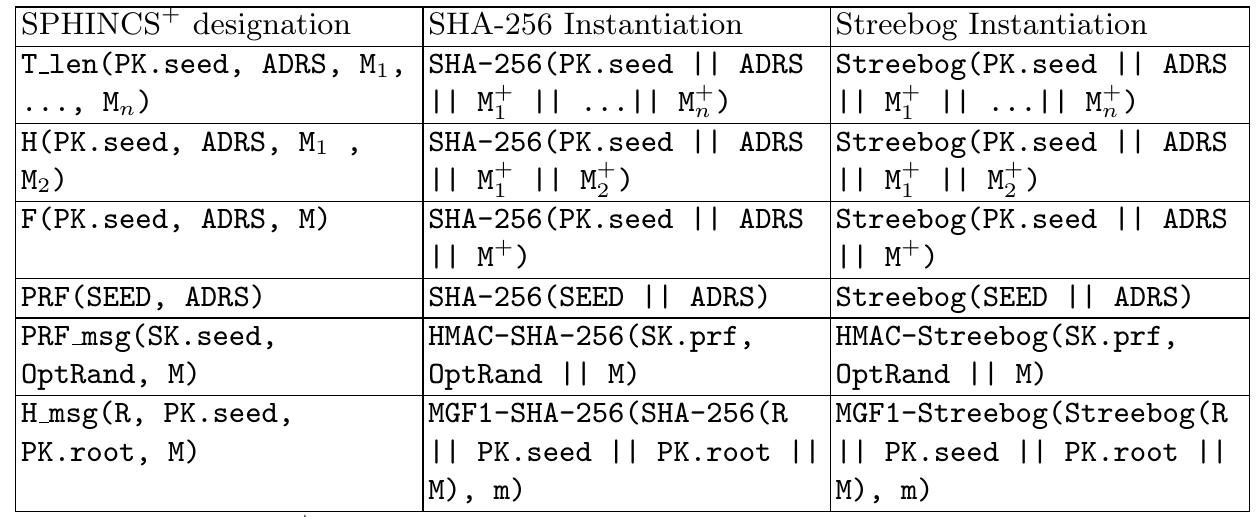}}
\centerline{\mbox{\textbf{Table.~III}. SPHINCS$^+$  primitives instantiation with SHA-256 and Streebog hash functions.}}
\label{table-inst}
\end{minipage}
\end{figure}

\subsection{SPHINCS$^+$ signature scheme} 

The SPHINCS$^+$ scheme uses a Merkle hyper-tree that consists of a large number of binary hash trees (see Fig.~1). 
Each node in the binary hash tree is associated with a hash string and has exactly two child nodes. 
The hash strings are computed as a function of children's strings. 
To build each subtree, several public-private key-pairs of Winternitz-type one-time signatures (WOTS$^+$) \cite{wots-plus} are generated.
Compressed public keys are then used as subtree leaves, whereas corresponding private keys sign roots of lower-level subtrees. 
On the lowest level, Winternitz keypairs sign public keys of FORS (Forest of Random Subsets) schemes. 
In their turn, FORS elements actually sign messages. 
For a given FORS element, the signature security decreases with every signed message. 
However, FORS elements in the hyper-tree are selected pseudorandomly, and probability to select the same FORS element multiple times degrades much faster. 
For this reason, there is no need to keep count of used FORS keys (``maintain the state"). 
The hyper-tree part follows closely Extended Merkle Signature Scheme (XMSS), which is a stateful signature scheme proposed for standardization by IETF~\cite{xmss-rfc}. 
Typically, there are 10 to 20 layers of subtrees with a height of each one about 3 to 8 depending on security level and other properties (see Table~I for various sets of parameters). 
The full hyper-tree is not held in memory rather the necessary parts are generated dynamically when signing takes place. 

A private key of the scheme consists of two seeds. 
The first one is used to pseudorandomly generate FORS and Winternitz secret key elements, and the other generates a randomization value for the message hashing. 
Public keys also have two elements: The root of the hyper-tree and a public seed that is used in various parts of the scheme. 
In practical implementations, the secret key holds copy of the public key elements. 

To sign a message, a randomizer is generated and the message is hashed with this randomizer. 
The output is split into message digest, and the part, which determines subtree and leaf position of a FORS element to use. 
The message digest is signed by the FORS scheme, and after that hyper-tree signature of the corresponding FORS public key is computed. 
It consists of Winternitz one-time signatures interleaved with Merkle authentication paths in subtrees. 
The full SPHINCS$^+$ signature includes FORS and hyper-tree signatures. 
For the verification purpose, FORS public key is reconstructed, then the hyper-tree root is computed and compared to the published value. 
The SPHINCS$^+$ security level (in bits), private key, public key and signature sizes (in bytes) for different specifications of the scheme are presented in Table~I according to Ref.~\cite{sphincs-pls}, 
even though the claimed security levels might be overestimated~\cite{sphincs-comments}.

The security of hash-based signature schemes can be reduced to underlying hash function properties. 
The SPHINCS$^+$ scheme can be built entirely from standard hash functions. 
However, it uses several auxiliary functions to wrap calls to them. 
They are summarized in Table~II. 

\subsection{SPHINCS$^+$ with Streebog Hash Function}

SPHINCS$^+$ describes instantiations of auxiliary functions in terms of three hash functions: SHAKE256, SHA-256, and Haraka. 
The Russian standardized hash function Streebog is of Merkle-Damgard type, which makes it similar to SHA-256. 
The compression function operates in the Miyaguchi--Preneel mode and employs a 12-round AES-like cipher.
Cryptoanalysis of Streebog hash function was a subject of intensive research~\cite{Wang2013,AlTawy2014,Ma2014}.

The necessary auxiliary functions and their instantiations in terms of SHA-256 and Streebog functions are summarized in the Table~III.
MGF1 is a hash-based mask generation function \cite{rfc2437} and HMAC is keyed-hashing for message authentication~\cite{rfc2104}
For SHA-256, we consider the function of the following form:
\begin{equation}
	M^+=M\oplus{\rm MGF1-\mbox{SHA-256}(PK.seed || ADRS, len(M))}.
\end{equation}
For the Streebog instantiations, it changes as follows:  
\begin{equation}
	M^+=M\oplus{\rm MGF1-Streebog(PK.seed || ADRS, len(M))}.
\end{equation}
In the SPHINCS$^+$ specification, three levels of security are considered. 
For each level, two sets of parameters are provided: one is optimized for speed and the other for signature size~\cite{sphincs-pls}. 

\begin{figure}[h]
\begin{minipage}[h]{1\linewidth}
\centerline{\includegraphics[width=0.7\columnwidth]{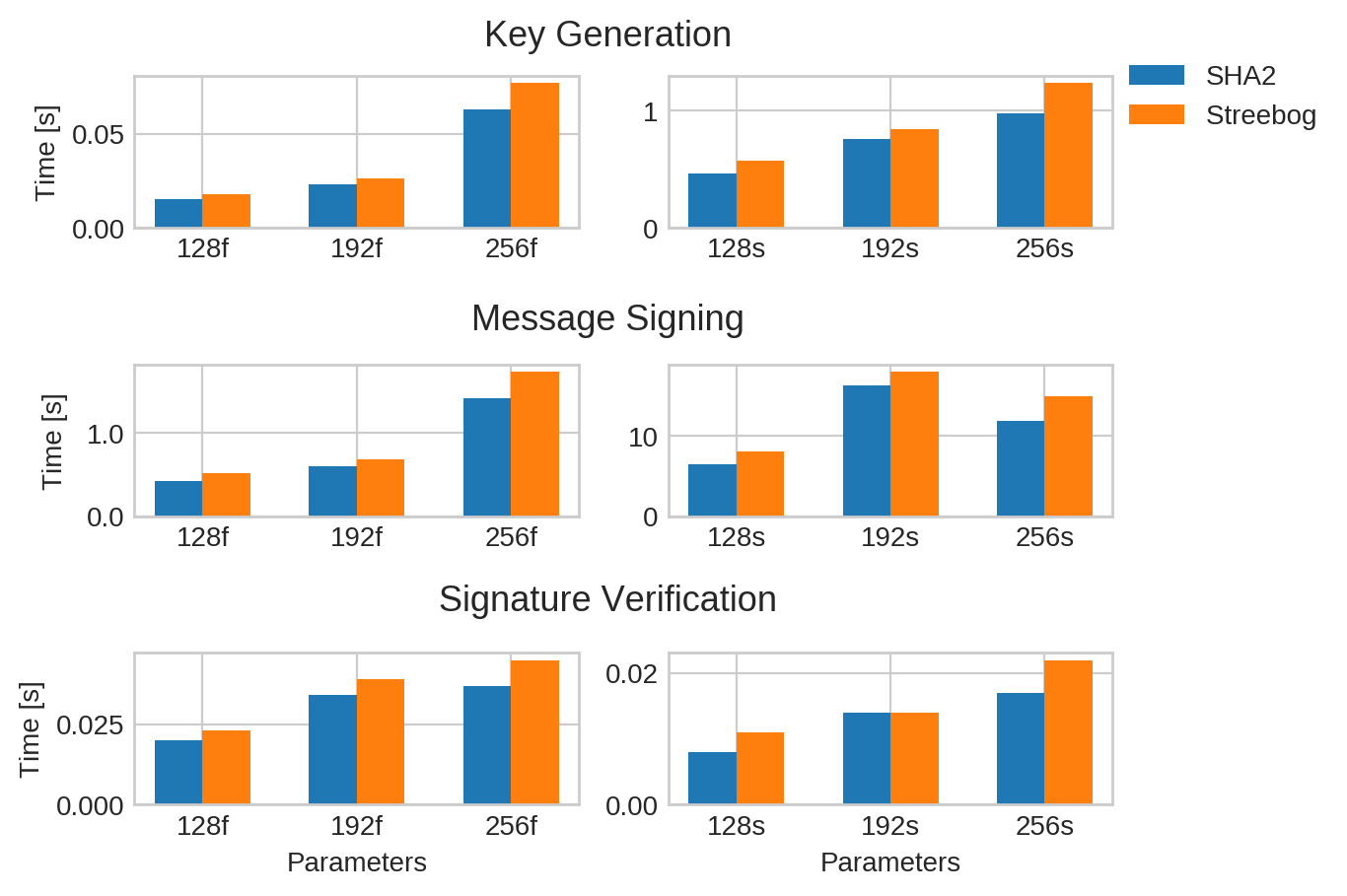}}
\centerline{\mbox{\textbf{Fig.~2}. Performance comparison of the SPHINCS$^+$-Streebog and SHA-256 instantiations for different sets of parameters.}}
\label{fig-benchmarks}
\end{minipage}
\end{figure}

We provide the results of the comparison of the performance the SPHINCS$^+$-Streebog and SHA-256 instantiations for each set of parameters. 
For the comparison purpose, we employ realizations of both hash functions from CryptoPro CSP 4.0.9958 Zhegalkin version library~\cite{CryptoPro}.
All the tests were performed on Xeon E5-2696v3 @ 2.3-3.8GHz processor with Linux 4.9 with the use of Google Benchmark Framework~\cite{Framework}.
The obtained results are illustrated in Fig.~2. 
One can see that for a particular implementation of the hash functions, Streebog achieves comparable performance, and thus is quite suitable for use in the SPHINCS$^+$ scheme.

\section{Conclusion}

We have analyzed how SPHINCS$^+$ hash-based digital signature scheme can be instantiated with hash function primitive Streebog, which is defined in GOST Russian Federation state standard. 
SPHINCS$^+$ scheme is provably secure, and its security depends only on properties of the underlying hash function. 
The Streebog hash function satisfies the demanded requirement for its use in the SPHINCS$^+$ digital signature scheme. 

\subsection*{Acknowledgments}

The work was partially supported by the Russian Foundation for Basic Research (18-37-20033). 
Authors would like to thank E.K. Alekseev, L.R. Akhmetzyanova, and L.A. Sonina for fruitful discussion and technical support in Streebog and SHA-256 implementations.

\end{document}